\documentclass[preprint,5p,times,twocolumn]{elsarticle}
\journal{Nuclear Instrumentation Method A}
\usepackage{amssymb}
\usepackage{siunitx}
\usepackage{lineno}
\usepackage{xspace}
\usepackage{floatrow}
\usepackage{hyperref}
\usepackage{natbib}
\usepackage{subfig}
\usepackage{upgreek}
\usepackage{caption}
\usepackage{caption3}
\usepackage{chemformula}

\def\IV {I--V\xspace}
\def \pn{p--n\xspace}

\def\drift     {\raisebox{.5pt}{\textcircled{\raisebox{-.9pt} {1}}}\xspace}
\def\induction {\raisebox{.5pt}{\textcircled{\raisebox{-.9pt} {2}}}\xspace}

\usepackage[version=4]{mhchem}
\usepackage{chemist}
\usepackage{mathtools}

\DeclarePairedDelimiterXPP\BigOSI[2]%
  {\mathcal{O}}{(}{)}{}%
  {\SI{#1}{#2}}
\newcommand\reaction[1]{\begin{equation}\ce{#1}\end{equation}}
\newcommand\reactionnonumber[1]%

\begin{document}
\begin{frontmatter}
\title{Fabrication of a Silicon Electron Multiplier sensor using \\ Metal Assisted Chemical Etching and its characterisation \tnoteref{t1}}
\author[a,b]{Marius M\ae hlum Halvorsen\corref{cor1}}
\author[a]{Victor Coco}
%\author[a]{Evangelos Leonidas Gkougkousis}
\author[a]{Paula Collins}
\author[b]{Heidi Sandaker}
\author[c,d]{Lucia Romano}

\address[a]{{CERN EP-LBD}, {Esplanade des Particules 1}, {Meyrin}, {1211}, {Switzerland}}
\address[b]{{University of Oslo, Department of Physics}, {}, {Oslo}, {0315}, {Norway}}
\address[c]{{Institute for Biomedical Engineering, University and ETH Zürich}, {8092}, {Zürich}, {Switzerland}}
\address[d]{{Paul Scherrer Institute}, {Forschungsstrasse 111}, {CH-5232}, {Villigen}, {Switzerland}}

\cortext[cor1]{Corresponding author at: CERN, Esplanade des Particules 1,1217 Meyrin, Switzerland.
E-mail address: marius.maehlum.halvorsen@cern.ch}

\begin{abstract}

The Silicon Electron Multiplier (SiEM) sensor is a novel sensor concept that enables charge multiplication by high electric fields generated by embedded metal electrodes within the sensor bulk. Metal assisted chemical etching (MacEtch) in gas phase with platinum as a catalyst has been used to fabricate test structures consisting of vertically aligned silicon pillars and strips on top of a silicon bulk. The pillars exceed \SI{10}{\micro \meter} in height with a diameter of \SI{1.0}{\micro \meter} and are arranged as a hexagonal lattice with a pitch of \SI{1.5}{\micro \meter}. Electrical characterisations through current -- voltage measurements inside a scanning electron microscope and a climate chamber have demonstrated that the MacEtch process is compatible with active media and \pn junctions.

\end{abstract}
\begin{keyword}
Silicon sensors\sep Timing detectors \sep Silicon electron multiplier \sep Charge multiplication \sep Metal assisted chemical etching
\end{keyword}
\end{frontmatter}

\section{Introduction}
\label{introduction}
The next generation of innermost tracker detectors will need sensors that can cope with extreme radiation environments. Both the instantaneous and integrated luminosity will increase. Sensors in an FCC-hh environment will demand radiation hardness approaching \SI{1e18}{1\, \MeV\, n_{eq}cm^{-2}}, while needing to provide and maintain a time resolution of tens of pico seconds \cite{ECFA}. Today's most prominent silicon sensor technologies fall into the categories of planar sensors, Low Gain Avalanche Diodes (LGADs) and 3D sensors, which can all in principle provide a sufficient time resolution. Planar sensors have demonstrated operation and signal extraction at a fluence of \SI{1.6E17}{1\, \MeV \, n_{eq}cm^{-2}}, however the collected charge is significantly reduced due to trapping \cite{Kramberger_2013}. Time resolution below $\SI{100}{\pico \second}$ is challenging to exploit using low power readout electronics due to the low signal and consequently relatively large jitter \cite{planar_time_res, NA62_stating_150ps}. The 3D sensors have a time resolution in the order of some tens of pico seconds as they benefit both from the large and fast signal due to the large charge generating volume and small drift distances, but have, however, a high capacitance and reduced fill factor for perpendicular tracks \cite{original_3D, timeres3d_zurich, 3D_KRAMBERGER201926}. Lastly, LGADs have an excellent time resolution, but their gain degrades after irradiation and diminishes towards \SI{2.5e15}{1\, \MeV \,  n_{eq} cm^{-2}}. For higher fluences, they display similar characteristics to planar detectors \cite{RADHARDLGAD, Gkougkousis_LGAD_COMP}.

The Silicon Electron Multiplier (SiEM) is a sensor concept that uses metal electrodes embedded within the silicon substrate. These electrodes can be biased to create high electric field regions in which charge multiplication occurs, providing large signals which can be read out with an excellent time resolution. Because the high field region is generated by metal electrodes rather than by doping, the gain is not expected to degrade by acceptor removal when the SiEM is irradiated. The bulk region where the charge is generated has no inefficient areas whatever the track angle.  The SiEM's expected performance and properties were studied with the means of simulation in ~\cite{HALVORSEN2022167325}.

\begin{figure}[h!]
    \centering
    \includegraphics[width=0.96\textwidth]{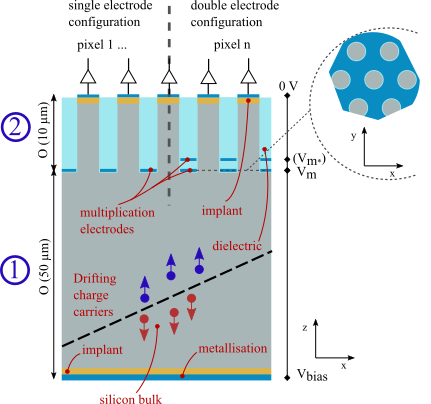}
    \caption{Schematic description of the SiEM sensor. It consists of a depleted silicon bulk region \drift adjacent to a region \induction made of silicon pillars (or strips). In between the pillars one or two metal electrodes can be implemented.
    The electrodes are biassed by $V_m$ and $V{_m*}$ and creates a high electric field region in \induction. When an ionising particle passes through \drift, primary electrons are generated and drift toward \induction where they get enough energy for impact ionisation and induce an amplified signal while drifting towards the readout electrode.}
    \label{fig:geometry}
\end{figure}

Figure~\ref{fig:geometry} illustrates the implementation of the concept. It comprises a silicon bulk region \drift and a region with silicon pillars and one or more multiplication electrodes \induction. 

Based on the simulations in \cite{HALVORSEN2022167325}, the pillars should have a height of \SI{\sim 10}{\micro \meter}, and be densely packed to minimize the charge carrier drift path inhomogeneities. This will also ease the full bulk depletion below the multiplication electrode. They should be thin, $\SI{\sim 2}{\micro \meter}$, for the electric field not to attenuate too much towards the center of the pillar. The multiplication electrodes can be seen in dark blue colors at the bottom of the trenches. They can be made with different geometrical properties and operated with different biasing configurations. Two geometrical configurations have been studied in simulations, with both single and double multiplication electrodes. 

In the single electrode configuration (Figure~\ref{fig:geometry}, left) the gain region is generated by the high field between the multiplication electrode biased at the potential $V_m$ and the readout electrode at ground. In the double electrode configuration (Figure~\ref{fig:geometry}, right) a second buried multiplication electrode is added and biased at $V_{m*}$. The high electric field region is in this case created in the pillar by the difference in potential between the two buried electrodes. According to TCAD simulations, gain is achievable in both configurations, but the field configurations differ such that the constraints on the materials are not the same. With the main voltage drop between $V_m$ and $V_{m*}$, a high field would be generated inside the dielectric, between the electrodes, and a milder field inside the silicon, while a single electrode configuration would be the opposite. An optimal geometry for a single electrode configuration would be densely arranged tall pillars. They should be tall in order to have a long high field region, which would yield more charge multiplication, as discussed in \cite{HALVORSEN2022167325}. However, they should not be too tall as the amount of silicon-oxide interface charge increases which degrades the device performance. A double electrode geometry can similarly be optimised with the separation of the multiplication electrodes, and should not be too separated to maintain the benefits of the field configuration.

Two methods are considered for fabrication: Metal Assisted Chemical Etching (MacEtch) and Deep Reactive Ion Etching (DRIE). The latter benefits from being a mature, industry standard technology, but has inherent defects such as ion induced defects and limitations due to scalloping, bottling, tapering, undercuts and aspect ratio dependent etching rate \cite{WU_HARE_review, campbell2008fabrication}. Metal Assisted Chemical Etching on the other hand is a rather recent technology, discovered in 2000 ~\cite{original_macetch}. It uses a metal catalyst to locally enhance the dissolution rate of silicon in the presence of an oxidant and an etchant. This accelerated etching gives an anisotropic pattern transfer that can yield very large aspect ratios (10 000:1) and small feature sizes ($\SI{\sim 10}{\nano \meter}$) ~\cite{gasetch}. 
The process has applications such as grating fabrication for X-ray optics \cite{gasetch}, photovoltaics \cite{macetch_photovoltaic}, creating vias \cite{vias_li}, and anode batteries \cite{macetch_batteries}. To the best of the authors' knowledge, MacEtch structures have not been used, until now, in active media with \pn junctions. \IV characteristics \cite{nanowires_electrical} and electrochemical properties \cite{nanowire_electrichemical} of MacEtch nanowires have been reported, indicating the phenomenon of charge carrier trapping at surface states. 

The MacEtch process can be suitable for the SiEM not only because of the small feature sizes, but also due to the etching catalyst serving as a multiplication electrode during the electrical operation of the device. MacEtch could thus reduce the amount of processing steps, and would particularly be suitable for a single electrode configuration. In this paper the fabrication and characterisation of a SiEM demonstrator made with the MacEtch process is described.

\section{Metal Assisted Chemical Etching for SiEM}
\label{section:MacetchForSiem}
In a MacEtch process, a metal mask, typically consisting of Au, Pt or Pd, is patterned onto a semiconductor substrate (Si) and works as catalyst for a local redox reaction occurring in presence of an acid etchant (\ce{HF}) and an oxidant (\ce{H2O2} or \ce{O2}). On contact with the metal, the oxidant is reduced by injecting holes into the semiconductor substrate, changing the oxidation state of the substrate and enabling the subsequent removal of the substrate material by the etchant. 
The MacEtch of silicon can be realized using \ce{HF} in liquid~\cite{original_macetch} or in vapor~\cite{gasetch, programmable_macetch} and different oxidants. In our experiment, the sample is normally placed above an \ce{HF} containing solution, such that \ce{HF} evaporates from the liquid tank and the oxidant is supplied from the air flow on the sample. The following reactions occur:

\reaction{O2 + 4H^+ + 4e^- -> 2 H2O (at catalyst) \label{chemeq:oxygen}}

\reaction{Si + 4h^+ +6HF -> SiF^{2-}_6 +6H^+  (at silicon) \label{chemeq:sitofluoride}}
.

Silicon can be dissolved directly (Eq. \ref{chemeq:sitofluoride}) forming fluorides species that are released as \ce{SiF4} in gas phase or via an indirect path, forming \ce{SiO2} ~\cite{gasetch}. As the Si is removed beneath the catalyst, the metal pattern sinks and contacts unreacted material, continuing the reaction to form a negative image of the metal mask.

\begin{figure}[h!]
    \centering
    \includegraphics[width=0.96\textwidth]{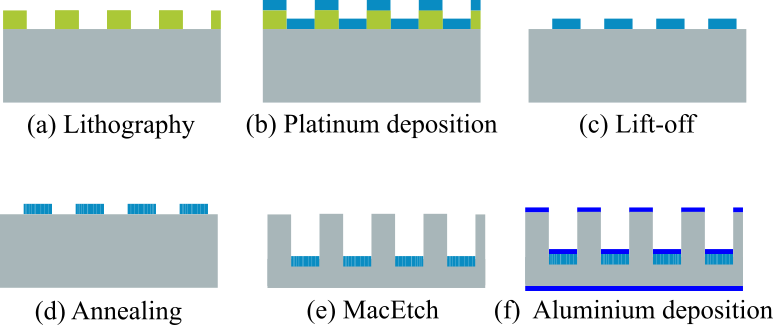}
    \caption{Overview of the main steps in the fabrication of the SiEM using MacEtch. }
    \label{fig:macetchschems}
\end{figure}

The main processing steps for the SiEM fabrication using MacEtch are summarised in Figure~\ref{fig:macetchschems}. The first step \textit{(a)} is the pattern application by UV-lithography. This is followed by a metal thin film evaporation \textit{(b)}. After lift-off \textit{(c)}, there is an annealing step to dewet and make the catalyst adhere to the silicon, \textit{(d)}. In step \textit{(e)} the sample is exposed to HF for the etching, and at the end of the fabrication, \textit{(f)}, a front and backside metallisation is performed for electrical contact.  

The fabrication was performed on a high resistivity (4--\SI{8}{\kilo\ohm\cm}) \SI{200}{\micro \meter} thick n-type (1~0~0) wafer already implanted with boron at the frontside to form the \pn junction and phosphorous at the backside for ohmic contact. A direct write laser lithography system was used to pattern two different structures \textit{(a)}: circles of \SI{1.0}{\micro \meter} diameter in an hexagonal lattice with a pitch of \SI{1.5}{\micro \meter} and strip-like geometries with the same pitch and width, but terminated with read-out pad allowing easier bonding and probing. The same processing steps were followed for both patterns. The dimensions were chosen to benefit from the MacEtch capabilities of small feature sizes and large aspect ratios, which would ease the depletion, give less lateral field attenuation inside the pillars and extended the high field regions.

\begin{figure}[!ht]
\centering
  \subfloat[]{\includegraphics[width=0.85\textwidth]{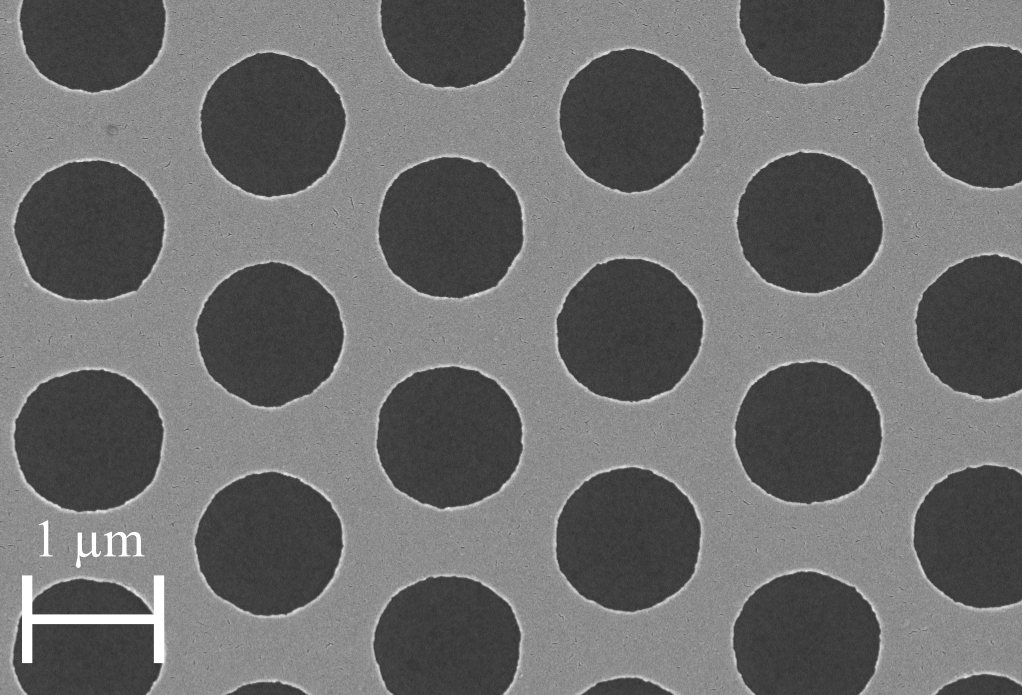}
  \label{fig:as_lift_off}
  }\\
  \subfloat[]{\includegraphics[width=0.819\textwidth]{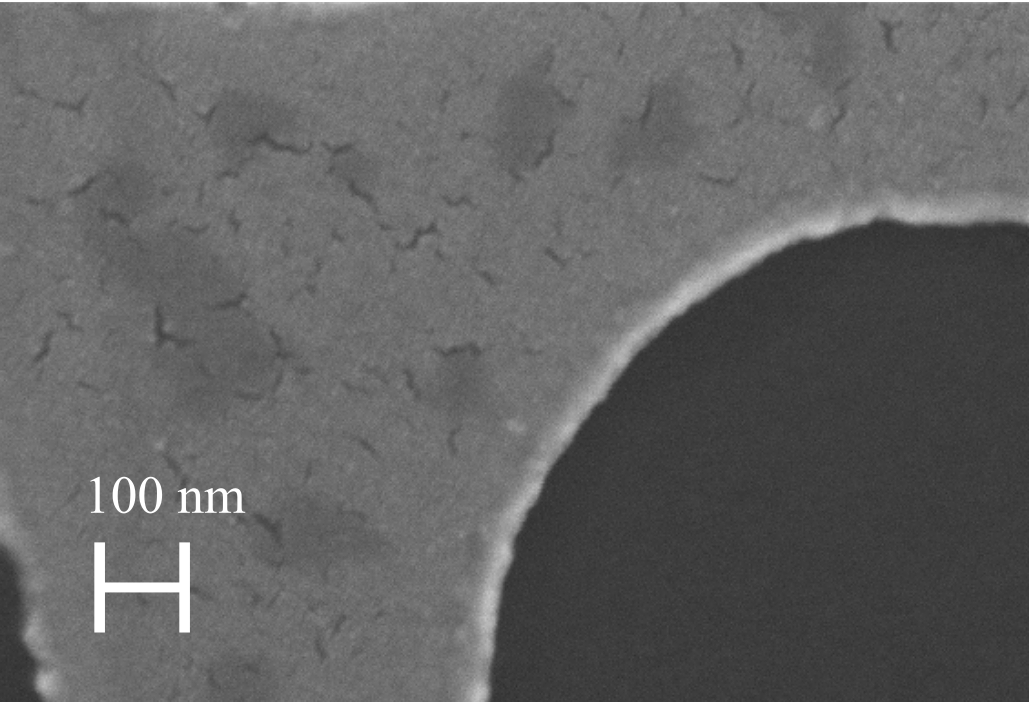}\label{fig:dewetted}}
  \caption{(a) Plan view of platinum patterned silicon sample after lift-off. (b) Nano pores due to self assembly of Pt.}
  \label{fig:before_etching}
\end{figure}

In this study \ce{Pt} has been chosen as a catalyst due to its excellent catalytic activity, and good adhesion due to the formation of silicides (\ce{PtSi} and \ce{Pt2Si}) at the interface. The silicides are formed by annealing at the interface after lift-off, step \textit{(c)} and \textit{(d)}. The thermal treatment also induceds the \ce{Pt} to self-assemble, creating nano-pores, a phenomenon referred to as dewetting \cite{Strobel_2010_dewetting}. The nano-pores enable a more uniform mass transport of the reactants across the pattern, which leads to a more uniform etching. Dewetting is preferred on an insulating substrate, therefore \ce{Pt} is evaporated onto the sample without the removal of the native oxide \cite{gasetch}. Scanning electron microscope (SEM) images of the sample after the Pt-deposition and lift-off can be seen in Figure \ref{fig:as_lift_off} and the nano pores after dewetting can be seen in Figure~\ref{fig:dewetted}.

\begin{figure}[h!]
    \centering
    \includegraphics[width=0.96\textwidth]{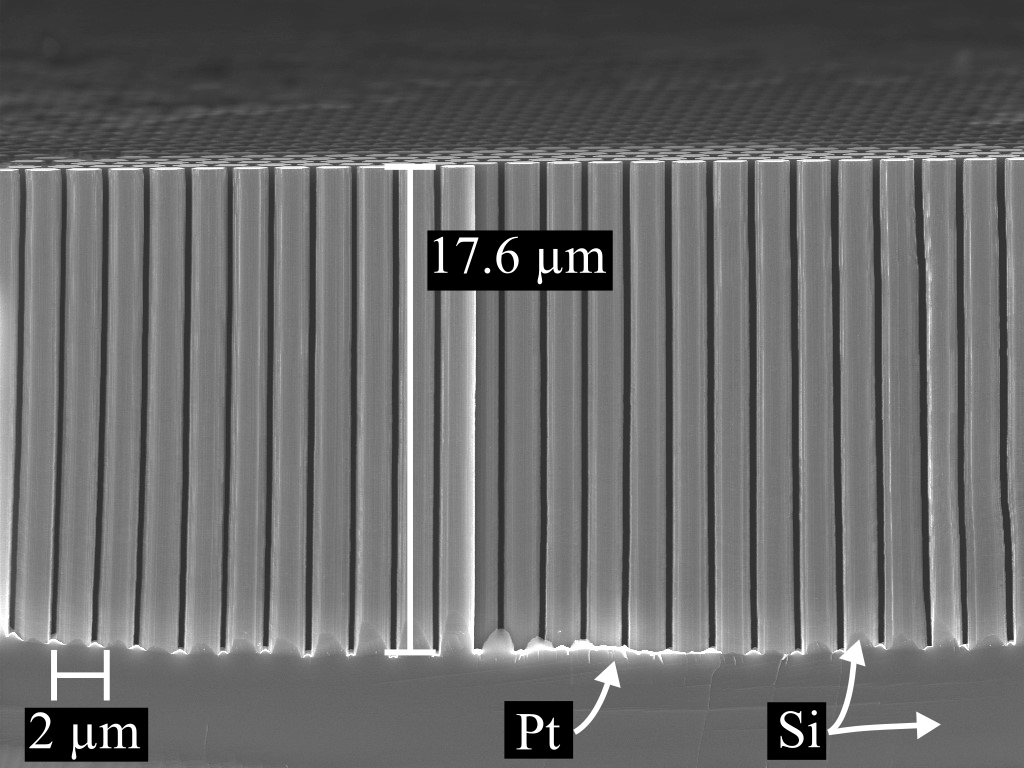}
    \caption{Cross section view of the silicon pillars after etching. The Pt thin film is visible at the bottom of the pillars.}
    \label{fig:etched_pillars}
\end{figure}

Previous work \cite{gasetch,liquid_etch} has demonstrated that geometries similar to what is needed for the SiEM can be produced both by etching in liquid and gas phase (step \textit{(e)}). 
While liquid etching is generally more prevalent, gas etching has certain advantages. In particular, it prevents the agglomeration of nano-wires, produced by the nano-pores mentioned before, which could lean onto the pillars and distort the fields, make shorts or undesired electrical states. In addition, the broad parameter space allows the etching quality to be enhanced by tuning the temperature, the flow of oxygen supply and the etching composition to optimise the etch-rate uniformity and - remarkably - avoid micro-porosity \cite{Hildreth_vapor}, which would introduce undesired defects in the device. Etching in gas state is thus chosen for this production. 
The sample was mounted on a heated sample holder and suspended above a HF solution. The temperature was kept at \SI{55}{\celsius}, which is found to be the best processing temperature considering etching uniformity, porosity reduction and agglomeration \cite{gasetch}. The resulting pillars after $\SI{\sim 1}{\hour}$ etching can be seen in Figure~\ref{fig:etched_pillars}. 

\begin{figure}[h!]
    \centering
    \includegraphics[width=0.96\textwidth]{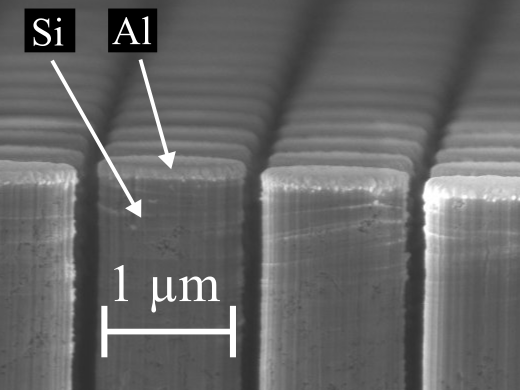}
    \caption{Cross section view of the top part of the pillar with aluminium deposited.
    }
    \label{fig:as_al_dep}
\end{figure}

For the sensor prototype a front and backside metallisation is needed in order to contact, deplete and read out signals.
It was achieved using evaporated aluminium, which can be seen in Figure~\ref{fig:as_al_dep}. Such an evaporation of aluminium may leave residuals of aluminium close to the top of the pillars, which could be problematic if conducting channels cross and shorten the \pn junction. For a demonstrator this approach is sufficient, but alternative approaches could be applied for later productions, which are further discussed in Section \ref{sec:future_work}. 
The native oxide was removed directly before the aluminium deposition in order to lower the metal--semiconductor interface barrier.

\section{Prototype details and electrical characterisation}

\begin{figure}[h!]
    \centering
    \includegraphics[width=0.96\textwidth]{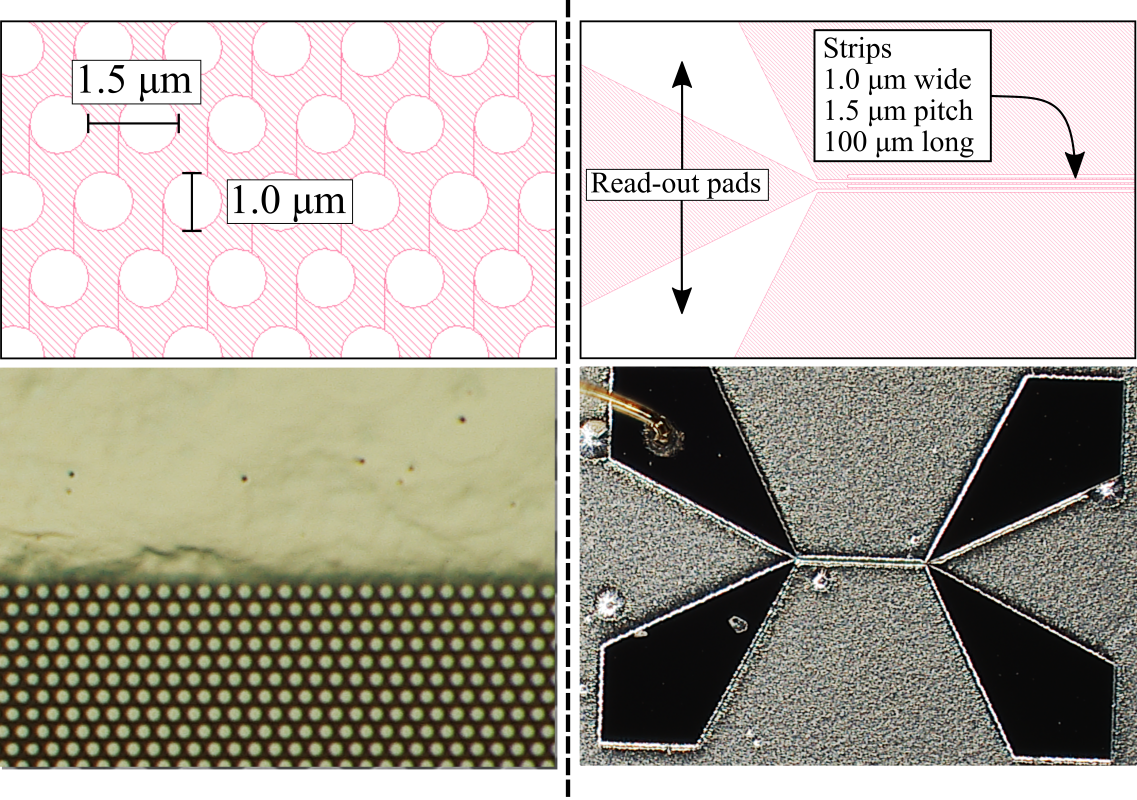}
    \caption{Snapshots of lithography mask and plan view images at optical microscope of the fabricated pillars (left) and strips (right). One of the strip readout-pads is bonded to a carrier board through gold ball wire bonding.}
    \label{fig:pattern_gds_planview}
\end{figure}

\begin{figure}[h!]
    \centering
    \includegraphics[width=0.96\textwidth]{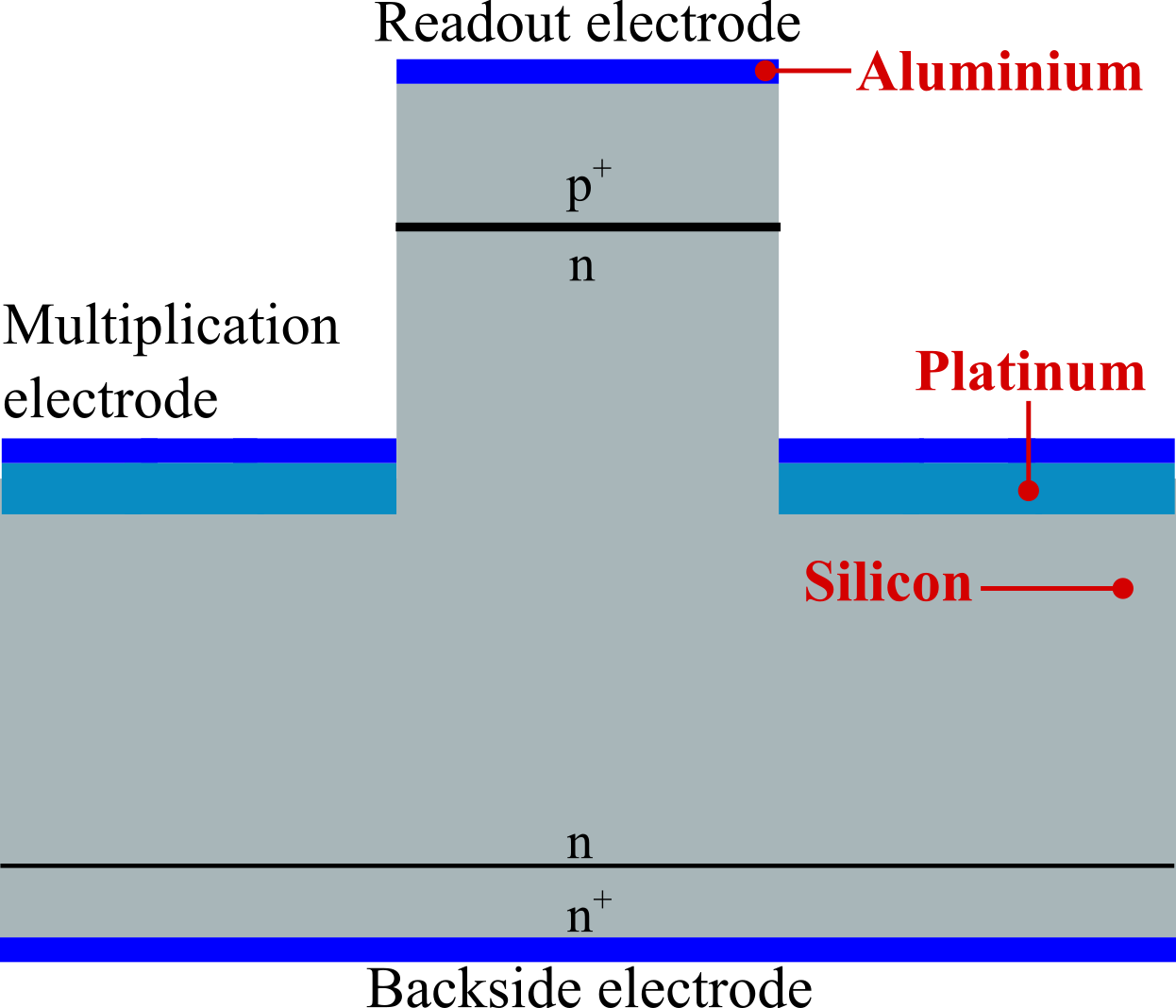}
    \caption{Schematic illustration of the prototype with the material type and contacts described. Dimensions are not to scale.}
    \label{fig:prototype_contacts}
\end{figure}

Images of the fabricated strips and pillars along with their patterns can be seen in Figure~\ref{fig:pattern_gds_planview}. The prototypes contain three electrical contact points: the readout electrode, the multiplication electrode and the backside electrode, mapping the single electrode configuration of Figure~\ref{fig:geometry}. The electrodes, doping and material composition of the prototype are illustrated in Figure~\ref{fig:prototype_contacts}.
The readout and backside electrodes are deposited directly on the $p^+$ and $n^+$ implants, giving ohmic contacts. The multiplication electrodes consist of the platinum catalyst and evaporated aluminium. \ce{Pt} and also \ce{Al} in direct contact with high resistivity n-type silicon are known to make a rectifying contact, a Schottky contact \cite{sze2006physics}.

Two setups have been used to assess the electrical properties of the structures through their current--voltage (\IV) characteristic. A probe system installed within a scanning electron microscope (SEM) was used after production of the sample. It allowed both the strips and pillar structures to be characterised. The strip structures could be further investigated in an \IV setup installed in a climate chamber after the structure had been connected to a carrier board with gold ball bonding. Measurements from these two setups are presented in the following.

\subsection{Measurements within SEM probe station}
\label{sec:SEM_IV}

An SEM Zeiss Supra VP55 equipped with Kleindiek Nanotechnik micro-manipulators with sub-micrometer tungsten needles and a Keithley 236 source measure unit was used to probe the \IV characteristics of single pillars and of the strip structures.

The sample was mounted on a glass plate to electrically isolate the sensor. The needles were lowered in several steps during which the focus was done successively on the needle tips and on the sample until contact with the sample was achieved. One needle was used to contact the readout electrode and one to contact the multiplication electrode. Once the needles were in place, the electron beam of the SEM was turned off in order to not interfere with the electrical characterisation of the structures. This biasing configuration was used to probe the multiplication region, \induction from Figure~\ref{fig:geometry}.

\begin{figure}[!ht]
\centering
  \subfloat[]{\includegraphics[width=0.81\textwidth]{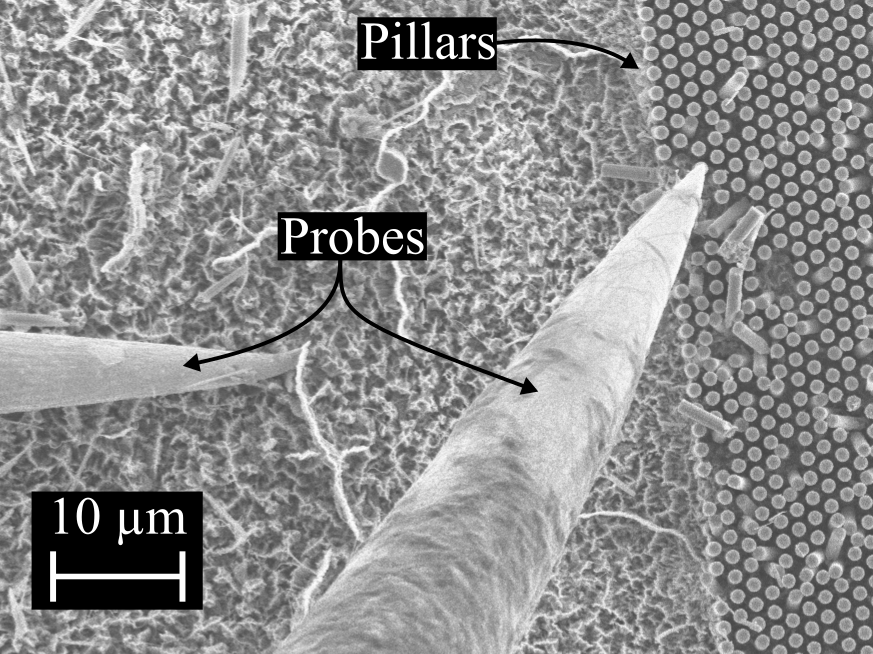}\label{fig:sem_single_electrode}}\\
  \subfloat[]{\includegraphics[width=0.81\textwidth]{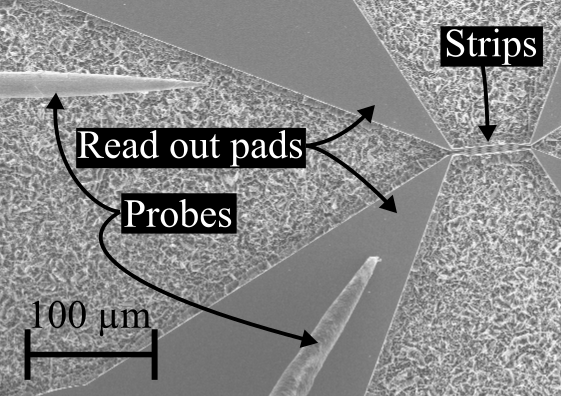}\label{fig:sem_strip}}\\
  \subfloat[]{\includegraphics[width=0.8\textwidth]{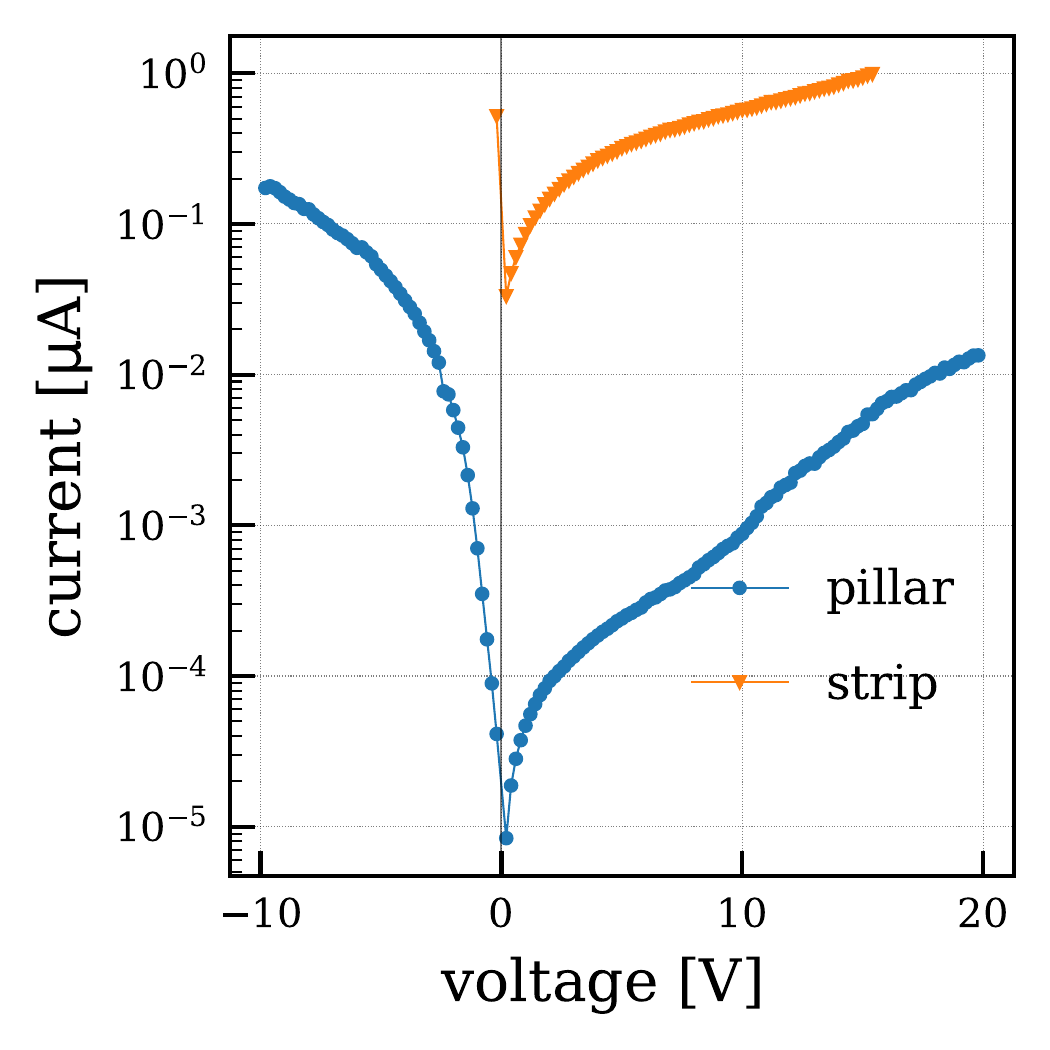}\label{fig:sem_IV}}

  \caption{Measurements with probe station inside a scanning electron microscope. The images are taken just before the voltage scan. \protect\subref{fig:sem_single_electrode} shows the contact with a single pillar, and \protect\subref{fig:sem_strip} shows the SEM image of the strip device. \protect\subref{fig:sem_IV} shows the \IV curve for both the strip and pillar. The strip curve stops at lower voltages because it reaches compliance of \SI{1}{\micro \ampere} quicker than the pillar. The positive output of the sourcemeter is applied on the multiplication electrode and a the negative on the readout electrode}
  \label{fig:sem_ivs}
\end{figure}

Figure \ref{fig:sem_single_electrode} shows the probing of exactly one single pillar right before the \IV measurements were taken. The strip probing is shown in Figure~\ref{fig:sem_strip}, where the needle was placed on the read-out pad of the strip. The \IV measurements of these two structures are shown in Figure \ref{fig:sem_IV}. The current of both structures display similar trends, with a slow rise towards positive voltages (reverse bias) and a steeper rise in current towards negative voltages (forward bias), typical for diodes.

However, the currents are rather large and it is not possible to identify a region of saturation current as in a standard diode. The current consists of several contributions, in particular the bulk generated, oxide interface generated and impact ionisation generated ones. The bulk generated and oxide generated currents mainly depend on the generation rates of the defects, their concentrations and the depletion volume for bulk generated and interface area for interface generated currents. The impact ionisation current depends mainly on the electric field. 

The rather high leakage current compared to conventional diodes \cite{RUZIN2002411} can be explained by the larger silicon--oxide interface areas and high defects concentration in these regions. These additional states at the interface would inject charge carriers into the system \cite{sze2001semiconductor}. 

The observed increase in leakage current for both structures is likely to originate from impact ionisation. This is because an increase in potential, with the given biasing scheme, would mainly enhance the field between the readout and multiplication electrode and not extend the depleted volume meaningfully. According to TCAD simulations the strips and pillars are mostly depleted even before biasing. 

Both desired and undesired impact ionisation can occur and contribute to the high leakage current. The latter can originate from imperfections in the pattern transfer, especially due to the dewetting that can give sharp edges and thus local high electric fields. Also the electrode edge in direct contact with the strips and pillars can give edge effects and avalanche regions in its proximity. 
The sharp edges could be smoothed by an additional oxidation step. It could also further separate the multiplication electrode from the pillar and strip which would reduce field extremities to give milder operation conditions inside the silicon.

Compared to the pillar, the strip gives a significant higher current, \SI{\sim e3} times the magnitude. It can be explained by the much larger feature sizes, substantially affected by the read-out pad (see Figure~\ref{fig:sem_strip}). However, the pure volumetric ratio of the strip and pillar is \SI{5.4e5}{}, which is larger than the measured difference. The oxide interface area ratio between the strip and pillar is \SI{2.1e3}{}, which corresponds more to the observed difference. This indicates that the surface generated current may dominate compared to the bulk. Further investigations with dedicated test structures are planned in order to better understand and separate the different current contributions.

\subsection{One dimensional \IV characterisation in climate chamber }
\begin{figure}[h!]
    \centering
    \subfloat[]{\includegraphics[width=8.cm]{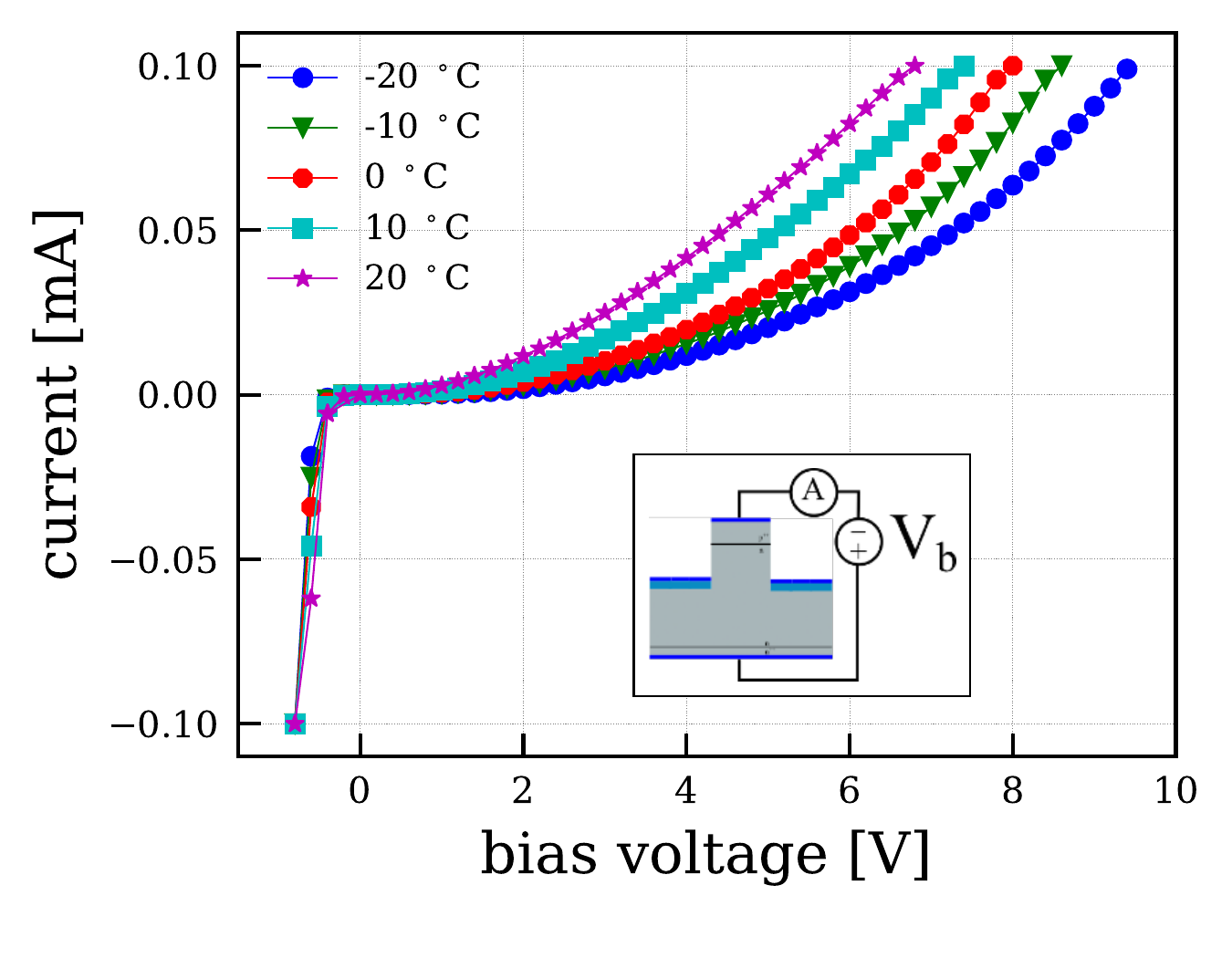}
        \label{fig:top_back_iv}    
    }\hfil 
    \subfloat[]{\includegraphics[width=8.cm]{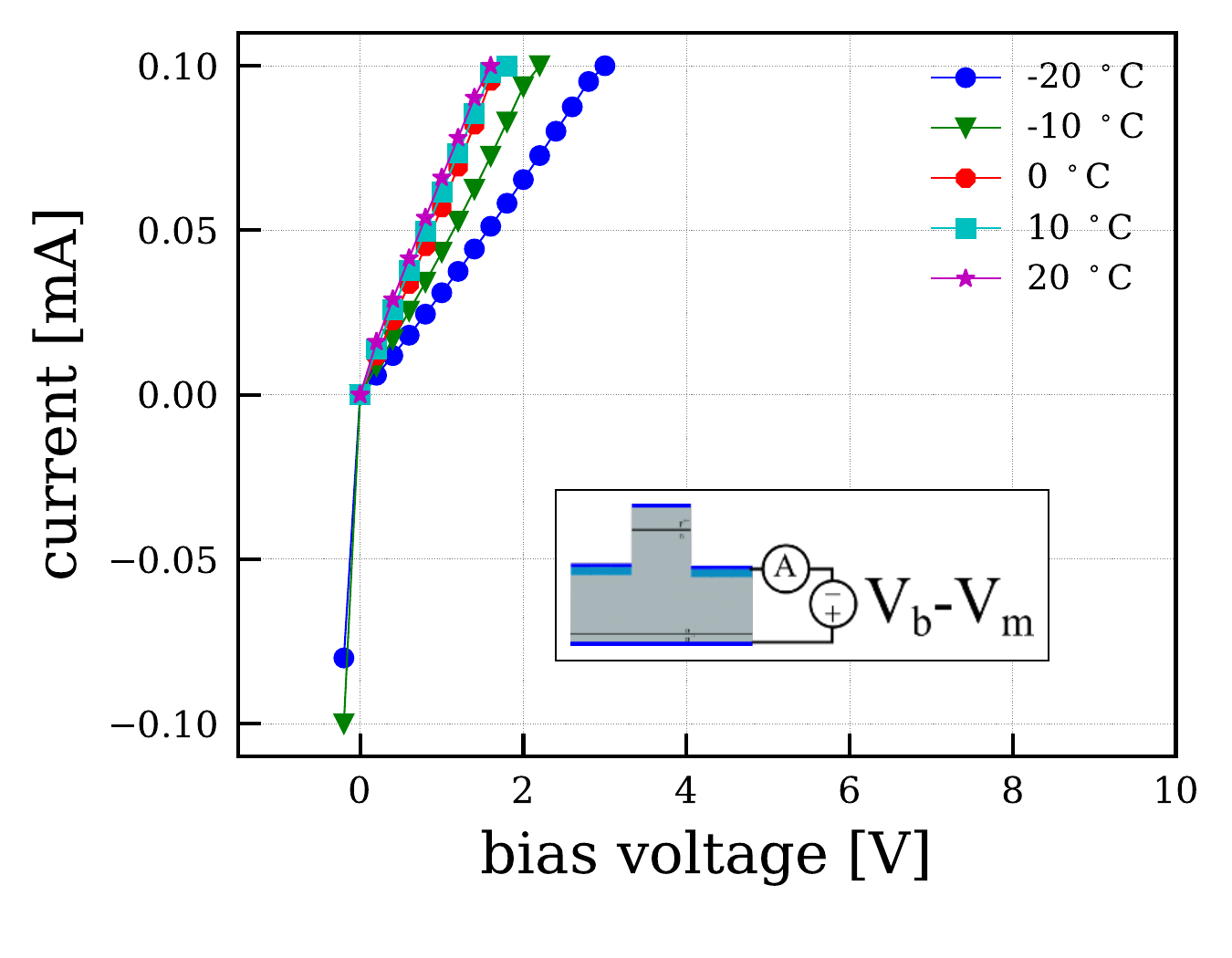}
        \label{fig:multi_back_iv}
    }\hfil
    \subfloat[]{\includegraphics[width=8.cm]{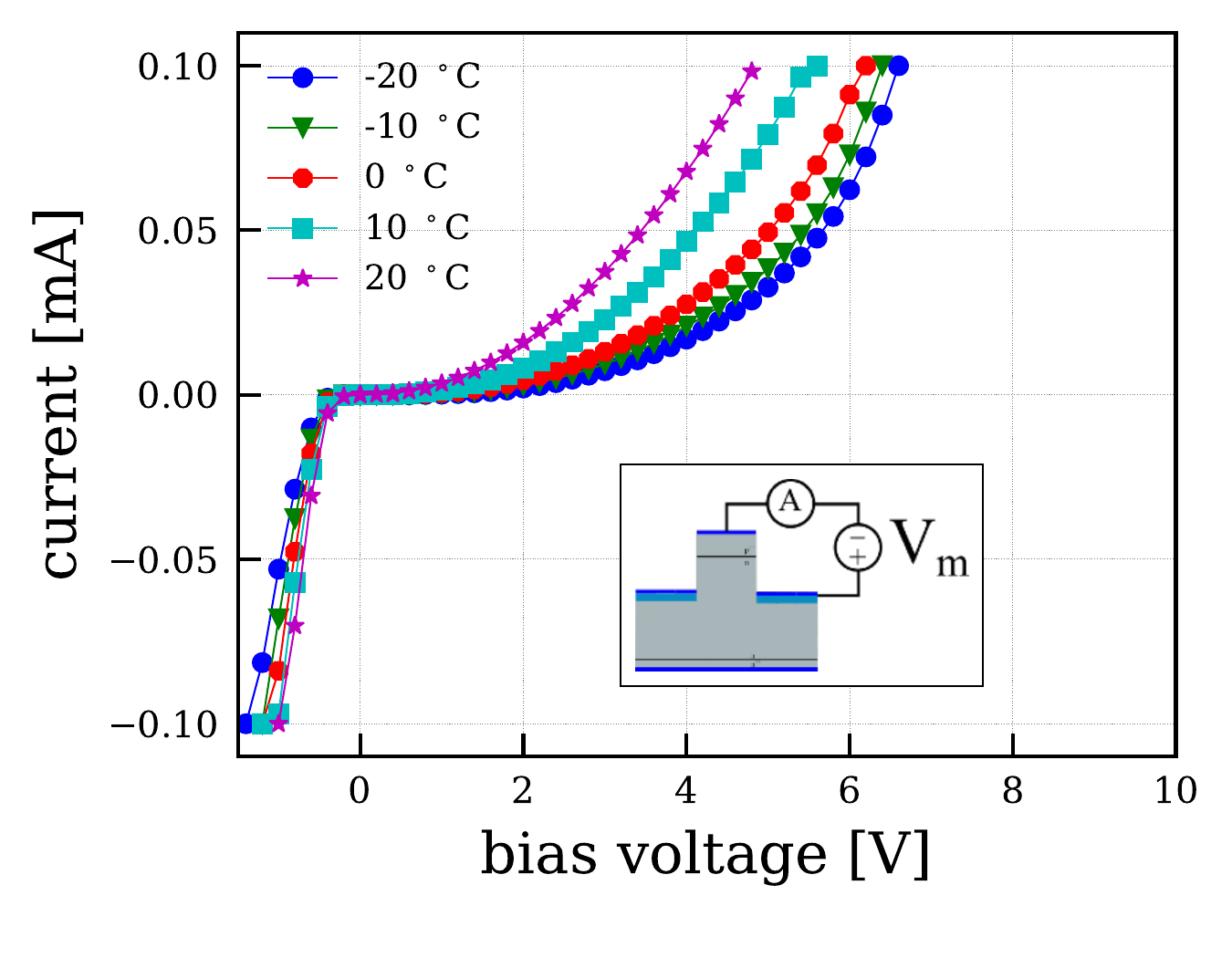}
        \label{fig:top_multi_iv}
    }
    \caption{\IV measurements between pairs of contacts at different temperatures. \protect\subref{fig:top_back_iv}~Shows the \IV when biasing the readout and backside electrode. \protect\subref{fig:multi_back_iv}~Shows the \IV when biasing the multiplication and backside electrode. \protect\subref{fig:top_multi_iv}~Shows the \IV when biasing the readout and multiplication electrode. All measurement points until the ampermeter maximum current setting are displayed.}
    \label{fig:all_1d_ivs}
\end{figure}

For the measurements in the climate chamber the readout pads of the strips were bonded to a carrier board by gold ball wire bonding, see right part in Figure~\ref{fig:pattern_gds_planview}. Their \IV characteristics were measured inside a climate chamber at five different temperatures, ranging from \SIrange{-20}{20}{\celsius}.
In this section, the \IV measurements of every possible two contact configuration, with the third contact left floating, are discussed. The measurements were taken using a Keithley 2410 sourcemeter and are shown in Figure \ref{fig:all_1d_ivs}. 

Figure \ref{fig:top_back_iv} shows the \IV between the readout electrode and backside electrode. It is expected to behave as a \pn junction and indeed it shows a much quicker increase in current towards negative voltages than positive voltages, typical of the \pn junction. It also shows a reduction of current with temperature, which is expected due to the lower generation rates.

Figure \ref{fig:multi_back_iv} shows the measurements while the multiplication and backside electrodes are biased. It is expected that in this case the currents display the characteristic of a Schottky diode, and indeed rectifying behaviour can be seen in the Figure \ref{fig:multi_back_iv}. In addition, compared to the \IV of the \pn junction discussed above, this current rises much more quickly. This is attributed to both the fact the electrode area is larger by a factor of around $\SI{2e3}{}$, and the fact the Schottky diodes inherently have larger leakage currents.

The last \IV characteristic measured is performed between the readout and multiplication electrode, thus the same as the one performed in the SEM. An equivalent representation of this system would be a \pn junction in series with a Schottky contact with the opposing polarity. The response and schematics of the system can be seen in Figure \ref{fig:top_multi_iv}. Again, the diode characteristic is clear. Compared to the \pn junction measurement using the readout and backside electrode in Figure~\ref{fig:top_back_iv}, the reverse bias currents rises faster and reaches the sourcemeter compliance of \SI{100}{\micro \ampere} earlier. This can be attributed to the fact that in this structure the electric field mainly develops in the strip, between the readout and the multiplication electrodes. Thus the possible depth of the depleted region is 30 times smaller than when the readout and backside electrodes are biased. The field magnitude is thus expected to be higher, yielding more impact ionisation and thus larger current. 

From the rectifying direction it can be seen that the \pn diode dominates the current characteristics. The Schottky diode is a majority carrier device whose forward \IV characteristic is mainly governed by thermionic emissions of electrons from the silicon conduction band to the metal. As most of the free electrons in the conduction band are evacuated due to the depletion, the Schottky forward current is limited by the reverse leakage current of the \pn junction. 

It can be seen that towards negative voltages, the current drops are less abrupt than in figure \ref{fig:top_back_iv}. This feature can be explained by the rectifying configuration of the Schottky contact, as it is in reverse mode once the \pn junction is in forward.

Compared to the measurements done in the SEM, the currents are substantially larger. Several factors can be influencing this difference. The SEM measurements were performed within three days after fabrication, while the ones measured in the climate chamber were measured a few months after production. During this time, growth of oxides and accumulation of defects at the interface can have occurred. In addition the gold ball wire bonding and mounting of sample onto the carrier board required some additional treatment steps (heating, dicing) that were not necessary in the SEM measurements. The samples originated from two different batches, made a few days apart, and small differences in the quality of, for instance, the metallisation and etching, can have occurred.

\subsection{Two dimensional \IV characterisation inside climate chamber }
The SiEM sensor with a single amplification electrode is meant to be operated by choosing the backside electrode bias $V_b$ such that the structure is depleted and the multiplication electrode bias $V_m$ such that the field is large enough to generate impact ionisation and thus amplify the signal charge. Hence all three electrodes must be contacted for the operation of the device. The interplay and current response when all the three electrodes are contacted and biased must therefore be understood. To do so $V_m$ and $V_b$ are varied in a "2D" \IV characterisation with the readout electrode kept at ground potential.

\begin{figure}[h!]
    \centering
    \includegraphics[width=0.96\textwidth]{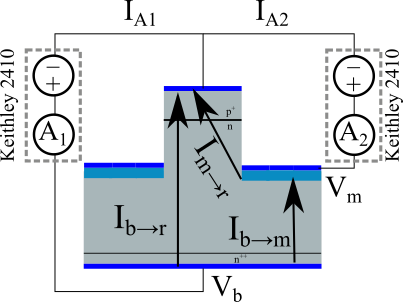}
    \caption{Illustration of the two dimensional \IV measurements, where two Keithley 2410 sourcemeters are connected together at ground on the readout electrode. One sourcemeter is connected to the multiplication electrode and the other to the backside electrode. The three current components are also indicated in the figure. }
    \label{fig:2D_iv}
\end{figure}

Two Keithley 2410 sourcemeters were used to bias and measure the system. Their ground outputs were connected together to the readout electrode. One of the sourcemeters biases and measures the current $I_{A_1}$ at the backside electrode while the other biases the multiplication electrode and measures $I_{A_2}$. The schematics of the measurement, the potentials and the measured currents can be seen in Figure~\ref{fig:2D_iv}. The arrows in the figure indicate the different current components present in the system, where $I_{b\rightarrow m}$ designates the current that flows from the backside to the multiplication electrode, $I_{b\rightarrow r}$ is the current that flows from the backside electrode to the readout electrode and $I_{m\rightarrow r}$ that flows from the multiplication electrode to the readout electrode. 

By adding the two measured currents $I_{A_1}$ and $I_{A_2}$ one gets
\begin{equation}
    I_{A_1}+I_{A_2}=I_{m\rightarrow r}+I_{b\rightarrow r},
    \label{eq:added_currents}
\end{equation}
thus the total current flowing to the readout electrode, without the contributions from $I_{b\rightarrow m}$. 

The voltage scan is performed by setting the backside electrode to a potential $V_b$, and then scan $V_m$ from \SI{0}{\volt} to $V_b$. It is not scanned further as $V_b-V_m$ would be negative, which corresponds to forward mode of the pure Schottky diode, and an abrupt forward current as seen in Figure~\ref{fig:multi_back_iv}.
The measurements are performed inside a climate chamber at \SI{-20}{\celsius}.

\begin{figure}[h!]
    \centering

   \includegraphics[width=0.9\textwidth]{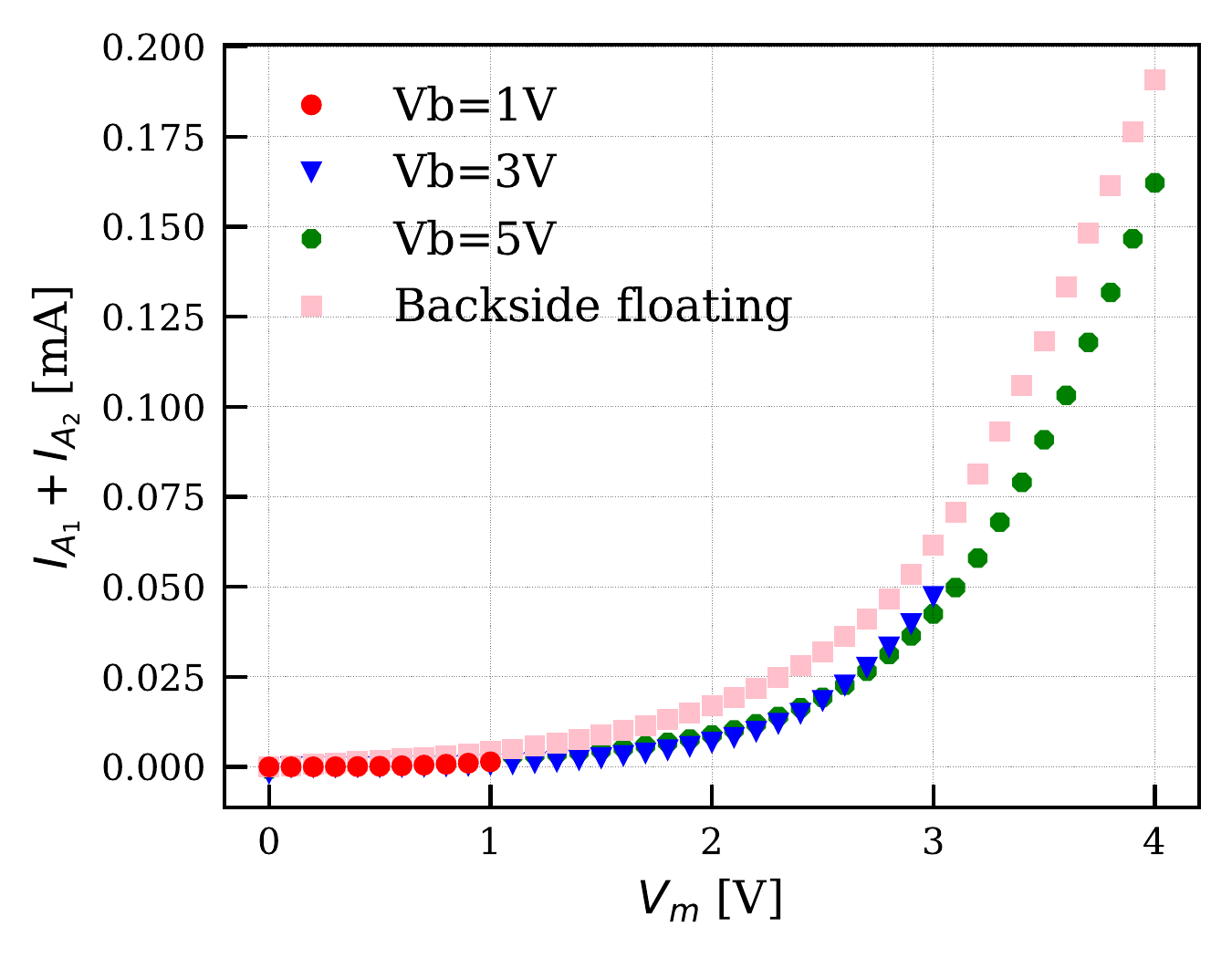}
    \caption{Two dimensional \IV measurements showing the total current through the readout electrode for a given $V_b$.}
    \label{fig:2D_iv_back_fixed}
\end{figure}

The total current through the readout electrode is plotted for the different bias configurations in Figure~\ref{fig:2D_iv_back_fixed}. 
The larger the bias voltage $V_{b}$ on the backside electrode, the larger is the range under which the multiplication electrode can be biased without a forward bias to the pure Schottky diode. Increasing $V_{m}$ increases
the amount of impact ionisation in the strip and thus the current increases, following the argumentation in Section \ref{sec:SEM_IV}. On the other hand increasing $V_{b}$ does not significantly increase the current as it mainly impacts the region \drift with relatively small fields. For all values of $V_b$, the total current at the readout is very close to the one when leaving the backside floating as in \ref{fig:top_multi_iv}. This shows further that the current at the readout electrode is dominated by the current in \induction which is generated by the amplification of the charges in the strip due to the large electric field and defects along the strip interface.

The summed 2D voltage scans show a lower leakage current than the 1D voltage scan with the backside floating. One possible reason could be field modifications around the multiplication electrode due to the contribution of the backside electrode. Instead of a rather abrupt transition from the depleted region to the neutral region at the multiplication electrode edge, the field can extend further into the bulk and potentially give milder working conditions by the multiplication electrode. However further investigations both through simulations and experiments should be performed to better understand this observed effect.

\section{Discussion and Future work}
\label{sec:future_work}
Results from the fabrication and characterisation of the Silicon Electron Multiplier sensor using Metal Assisted Chemical Etching have been presented demonstrating the fabrication approach along with the current response to different biasing configurations. Throughout the article possible shortcomings and new approaches have been introduced. This section aims to further discuss improvements and give an outlook for forthcoming studies of the SiEM.

\textit{Metallisation:}
A changed metallisation procedure could be implemented to avoid residual metal along the pillar wall that could potentially shorten the \pn junction. Approaches such as hot embossing of metal film \cite{hot_embossing} followed by a second lithography, or trench filling of dielectric before metal evaporation and a second lithography should limit this effect. These approaches would also allow several pillars to be grouped together, for instance to target a given readout ASIC pitch. 
Alternatively, a deep junction could be implemented into the starting wafer to ensure no metal channels cross the junction.

\textit{Etching:}
The silicon nano-wires, which result from the dewetting cracks in the Pt layer, can agglomerate when the aluminium is deposited. They can then lean onto the pillars, which could be a potential source of current noise. The nano-wires can however be minimised by increasing the Pt film thickness and etching in an oxygen rich environment, which should be used as a standard approach.

\textit{Oxidation steps:}
Sharp edges due to imperfections in pattern transfer and edge effects next to the multiplication electrode could be minimised by an oxidation step. This would smooth the interface, thus limiting both the high fields due to sharp edges and in addition separating the electrode from the etched out wall such that the fields and operation conditions in the silicon are milder.

\textit{Pattern:}
The first version of the pattern included a very large multiplication electrode. The extent of this electrode should be limited in future patterns to minimise the Schottky diode contribution. 

\textit{Material:}
In the first production a high resistivity n-type wafer was used. A second production should use a p-type wafer in order to collect and multiply electrons. 

\textit{Measurements:}
The next step of the electrical characterisation is to probe the sensor response to ionising radiation. Lasers and radioactive sources should be used to study pulse shapes, fill factor, collection efficiency and to make comparisons with simulations. A study should be performed to better understand the currents observed. Interface area scaling could be studied by using circular diode-like structures with different diameters. Defect spectroscopy, in particular between the multiplication electrode and readout electrode, should be performed to further understand the defects.

\textit{Other fabrication approaches:}
The production of a demonstrator using the Deep Reactive Ion Etching based process is currently ongoing. This approach has other constraints and benefits. There is no metal catalyst needed in the etching, which allows the introduction of a dielectric between the metal and silicon. Several electrodes can be introduced by consecutive deposition of metal and oxide layers. The same aspect ratios and feature sizes are however more challenging to achieve.

\section{Conclusion}
The first demonstrator of the Silicon Electron Multiplier sensor has been fabricated using metal assisted chemical etching (MacEtch). Electrical characterisations with two and three biasing points performed inside an SEM and a climate chamber have proven that the diode characteristics are preserved after processing, demonstrating the MacEtch compatibility with active media and a \pn junction. 
 
\section*{Author Statement}
\textbf{Marius M\ae hlum Halvorsen:}
Investigation, Methodology, Writing - Original Draft, Visualization, Data Curation, Formal analysis.

\textbf{Victor Coco:}
Supervision, Conceptualization, Funding acquisition, Writing - Review \& Editing

\textbf{Paula Collins:}
Funding acquisition, Writing - Review \& Editing

\textbf{Heidi Sandaker:}
Supervision

\textbf{Lucia Romano:}
Supervision, Methodology, Project administration, Investigation, Resources, Writing - Review \& Editing

\section*{Acknowledgment}
The authors thank the TOMCAT team with prof. M. Stampanoni (ETH Zurich) for facilitating the project at Paul Scherrer Institute; D. Marty, V. Guzenko, K. Vogelsang, K. Jefimovs for technical support in the PSI clean room facilities;  Z. Shi, C. Lawley for useful discussions and help in the fabrication; A. Weber and D. Kazazis for support at SEM probe station; S. Tsujino for the source measure unit; C. Scarcella and J. Alozy for support with Gold Ball Wire Bonding; J. Haimberger for help useful discussions, help with measurements and careful proofreading; V. Gkougkousis for measurement software; M. van Beuzekom for useful discussions; CERN QARTlab for optical microscope. This work was funded by and carried out in the context of the CERN Strategic R\&D Programme on Technologies for Future Experiment~\href{https://ep-rnd.web.cern.ch/}{https://ep-rnd.web.cern.ch/} and has received funding from the European Union’s Horizon 2020 Research and Innovation programme under GA no 101004761.

\bibliographystyle{elsarticle-num} 
\bibliography{main}

\end{document}